\documentclass[aps,prL,english,reprint,preprintnumbers,superscriptaddress,amsmath,amssymb,bibnotes,longbibliography]{revtex4-2}
\usepackage{graphicx}
\usepackage{dcolumn}
\usepackage{epstopdf}
\usepackage{bm}
\usepackage{flushend}

\usepackage[pdfstartview=FitH, CJKbookmarks=true, bookmarksnumbered=true, bookmarksopen=true, colorlinks, pdfborder=001, linkcolor=blue, anchorcolor=blue, citecolor=blue,urlcolor=blue,breaklinks]{hyperref}
\usepackage[mathlines]{lineno}

\begin{document}

\title{Phonons and magnetism of kagome FeGe probed by nuclear resonant scattering}

\author{Yu Tang}
\affiliation{Center for Correlated Matter and School of Physics, Zhejiang University, Hangzhou 310058, China}

\author{Saizheng Cao}
\affiliation{Center for Correlated Matter and School of Physics, Zhejiang University, Hangzhou 310058, China}

\author{Sijie Xu}
\affiliation{Department of Physics and Astronomy, Rice University, Houston, Texas 77005, USA}
\affiliation{Rice Laboratory for Emergent Magnetic Materials and Smalley-Curl Institute, Rice University, Houston, Texas 77005, USA}

\author{Xiaokun Teng}
\affiliation{Department of Physics and Astronomy, Rice University, Houston, Texas 77005, USA}

\author{Sven Velten}
\affiliation{Deutsches Elektronen-Synchrotron DESY, Notkestr. 85, 22607 Hamburg, Germany}

\author{Ilya Sergeev}
\affiliation{Deutsches Elektronen-Synchrotron DESY, Notkestr. 85, 22607 Hamburg, Germany}

\author{Pengcheng Dai}
\affiliation{Department of Physics and Astronomy, Rice University, Houston, Texas 77005, USA}
\affiliation{Rice Laboratory for Emergent Magnetic Materials and Smalley-Curl Institute, Rice University, Houston, Texas 77005, USA}

\author{Yilin Wang}
\affiliation{School of Emerging Technology, University of Science and Technology of China, Hefei 230026, China}
\affiliation{Hefei National Laboratory, Hefei 230088, China}
\affiliation{New Cornerstone Science Laboratory, University of Science and Technology of China, Hefei 230026, China}

\author{Yu Song}
\email{yusong\_phys@zju.edu.cn}
\affiliation{Center for Correlated Matter and School of Physics, Zhejiang University, Hangzhou 310058, China}

\begin{abstract}
Kagome FeGe hosts a $2\times2\times2$ charge-density wave (CDW) that strongly interplays with antiferromagnetic order. Here, we report $^{57}$Fe nuclear resonant scattering measurement to study FeGe across its long-range CDW and incommensurate magnetic transitions. Upon entering the CDW state, hardening of acoustic phonons and optical phonons around 22~meV, 27~meV, and 31~meV are observed in the Fe partial phonon density of states, which can be qualitatively captured by first-principle calculations. Upon entering the incommensurate magnetic phase, neither the phonon density of states nor the hyperfine interaction parameters change significantly, although a subtle feature associated with the incommensurate magnetic order or slow fluctuations is detected in the time-domain M\"{o}ssbauer spectra. These findings show that the CDW in kagome FeGe significantly modifies its lattice dynamics and magnetism,  evidencing an intertwined nature of the spin, charge, and lattice degrees of freedom.

\end{abstract}

\maketitle

\section{Introduction}

The interplay of geometric frustration, topological bands, van Hove singularities, and electronic correlations in kagome metals give rise to a wealth of emergent quantum phases, with entwined charge, spin, and lattice degrees of freedom. When the flat band or van Hove points of the kagome lattice are tuned close to the Fermi level, electronic correlations may become enhanced and stabilize unconventional phases of matter  \cite{Neupert2021,Yin2022,jiang_kagome_2023,Wang2023,Wilson2024,Zhou2024}. 

\begin{figure}
	\includegraphics[width=1\columnwidth]{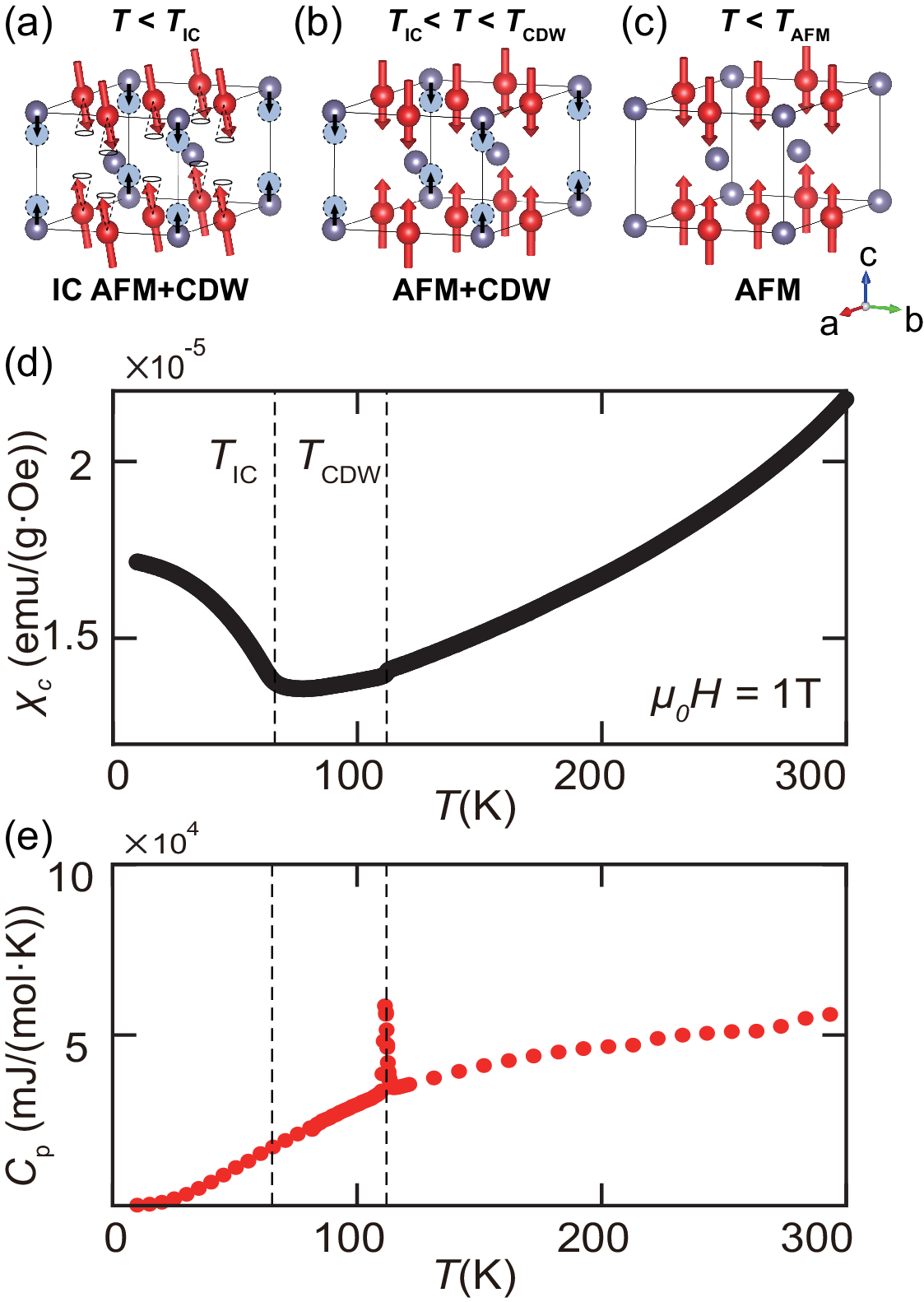} \protect\caption{Schematic crystal and magnetic structures of FeGe for (a) $T<T_{\rm IC}$, (b) $T_{\rm IC}<T<T_{\rm CDW}$, and (c) $T<T_{\rm N}$ \cite{chen_competing_2024}. (d) The magnetic susceptibility of FeGe measured in a 1~T $c$-axis field. (e) The specific heat of FeGe.}
	\label{Fig_basic}
\end{figure}

Recently, kagome FeGe is found to exhibit a $2\times2\times2$ charge-density-wave (CDW) that coexists with antiferromagnetic (AFM) order \cite{teng_discovery_2022,Yin2022PRL,teng_magnetism_2023,Yi2025}. Compared to its weakly-correlated vanadium-based counterparts \cite{Wilson2024,Zhou2024,Arachchige2022}, significant electronic correlations are present in FeGe, leading to a renormalized electronic structure \cite{teng_magnetism_2023} and magnetic spectral weight well beyond density functional theory \cite{chen_competing_2024}. A partial dimerization of Ge atoms situated within the Fe kagome plane is suggested to drive the CDW transition in FeGe [Fig.~\ref{Fig_basic}(b)] \cite{miao_signature_2023}, and leads to an enhanced spin polarization \cite{wang_enhanced_2023}. While a short-range CDW was initially reported \cite{teng_discovery_2022}, further studies found that annealing at 320~$^\circ$C induces a long-range CDW with an enhanced ordering temperature, implicating a role of Ge vacancy/disorder in the annealing-tuning of the CDW and magnetism of FeGe \cite{wu_annealing-tunable_2024,shi_annealing-induced_2024,chen_discovery_2024,Klemm2025}. 

Upon cooling, annealed FeGe exhibits $A$-type AFM order below $T_{\rm N}\approx400$~K, a long-range CDW via a first-order transition below $T_{\rm CDW}\approx110$~K, and an incommensurate (IC) magnetic order below $T_{\rm IC}\approx60$~K \cite{wu_annealing-tunable_2024,Klemm2025,Klemm2025b} [Figs.~\ref{Fig_basic}(a)-(c)]. The CDW phase results from a 1/4 dimerization of Ge atoms along the $c$-axis, which also induces in-plane charge modulations \cite{chen_discovery_2024}. The IC AFM state is characterized by a double cone magnetic structure, where the $c$-axis moments are modulated by ${\bf Q}=(0,0,\frac{1}{2})$ which corresponds to $A$-type AFM, and the in-plane moments are modulated by $(0,0,\epsilon)$, with $\epsilon\approx0.46$ \cite{Bernhard1988}. Subsequent inelastic neutron scattering measurements found incommensurate spin excitations above $T_{\rm IC}$, suggesting that the commensurate order of $c$-axis moments arise from localized electrons, whereas the incommensurate order of $ab$-plane moments are due to itinerant electrons \cite{chen_competing_2024,Klemm2025}. More recent angle-resolved photoemission spectroscopy and neutron scattering measurements evidence the IC magnetic order to be a spin-density wave state intertwined with the CDW \cite{Klemm2025b, Oh2025}.

The coexistence of AFM and CDW orders in FeGe allows for the coupling and interplay between charge, spin, and lattice degrees of freedom \cite{teng_discovery_2022,teng_spin-charge-lattice_2024}. The effect of the CDW and the IC magnetic orders on the lattice dynamics can be directly probed through the evolution of phonon density of states (PDOS) across $T_{\rm CDW}$ and $T_{\rm IC}$. Given the prominent role of disorder and the tendency towards phase separation \cite{Tan2025,Klemm2025}, the evolution of magnetism in FeGe should also be investigated via local probes, complementing neutron scattering measurements that probe the bulk average.

In this work, we use synchrotron nuclear inelastic scattering (NIS) to probe the Fe partial PDOS and time-domain M\"{o}ssbauer spectroscopy to study the hyperfine interactions in annealed FeGe, across both $T_{\rm CDW}$ and $T_{\rm IC}$. The Fe partial PDOS exhibits significant changes across $T_{\rm CDW}$, with significant hardening occurring for acoustic phonons, and optical phonons around 22~meV, 27~meV, and 31~meV. Comparison with first-principles calculations shows that these changes across $T_{\rm CDW}$ are qualitatively reproduced.
In contrast, the Fe partial PDOS hardly changes across $T_{\rm IC}$, indicating a weak coupling of the IC magnetic order to the lattice. The time-domain M\"{o}ssbauer spectra reveal a clear increase in magnetic hyperfine field below $T_{\rm CDW}$, evidencing enhanced AFM order in the CDW state via a local probe. Increase in the time spectra intensity around $t\approx65$~ns and $\approx85$~ns occur below $\approx70$~K, and can be attributed to the onset of IC magnetic order or corresponding slow fluctuations. Concomitantly, the hyperfine field hardly changes, suggesting that the total ordered moment in the IC state remains similar to values above $T_{\rm IC}$. 

\section{Methods}
\subsection{Experimental details}
Single crystals of FeGe were grown using the chemical vapor transport method and then annealed in vacuum at 320°C for 96 hours \cite{Klemm2025}. Magnetic susceptibility and specific heat measurements were conducted using a Quantum Design Physical Property Measurement System (PPMS-9T). The $^{57}$Fe NIS and time-domain M\"{o}ssbauer (nuclear forward scattering) measurements were performed at the High Resolution Dynamics beamline P01 \cite{Wille2010} of PETRA~III, DESY. The storage ring was operated in timing mode with 40 bunches separated by 192~ns. The X-rays were monochromatized to an energy bandwidth of 1.0 meV around the $^{57}$Fe nuclear resonance at 14.413~keV using a high-resolution monochromator (HRM). The nuclear resonance signal was separated in time from electronic scattering and measured using Si avalanche photodiodes with a time resolution of about 1~ns. The samples were mounted into a He closed-cycle cryostat with a temperature range of 10~K - 300~K. The NIS spectra were measured by tuning the HRM energy in 0.2~meV steps over the range $-20$~meV to $+80$~meV around the nuclear resonance energy. The real sample temperature was determined from the detailed thermal balance of the negative and positive parts of the NIS spectra. Based on these sample temperatures (44~K, 104~K, 128~K, 298~K) and their differences from measured cryostat temperatures, the sample temperatures for time-domain M\"{o}ssbauer measurements were estimated. The $^{57}$Fe partial PDOS were evaluated from the NIS spectra using the procedure described in Ref.~\cite{Kohn2000}.

\subsection{First-principles phonon calculations} 
The DFT calculations are performed using the \texttt{VASP} package \cite{VASP1,VASP2}, with the generalized gradient approximation (GGA) exchange-correlation functional \cite{GGA}. The experimental lattice parameters $a=4.985$~{\AA} and $c=4.048$~{\AA} are used \cite{teng_discovery_2022}. Energy cutoff of the plane-wave basis is set to be 500~eV. The criterion of total energy convergence is set to be 10$^{-8}$~eV, and the internal atomic positions are relaxed until the force of each atom is smaller than 1~meV/{\AA}. Phonon calculations are performed in the A-type AFM phase using the supercell (frozen phonon) method, with the \texttt{Phonopy} package \cite{TOGO20151}. $2\times2\times2$ and $4\times4\times4$ supercells (with respect to the nonmagnetic $1\times1\times1$ cell) are used for the pristine and CDW states, respectively. $\Gamma$-centered $k$-point grids of $16\times16\times10$ and $8\times8\times10$ are used for the charge-self-consistent calculations, and $8\times8\times10$ and $4\times4\times5$ $K$-point grids are used for phonon calculations, for the pristine and CDW states, respectively. For the calculations of phonon density of states, dense $q$-point grids of $41\times41\times21$ and $21\times21\times21$ are used for the pristine and CDW states, respectively.

\section{Results}
\subsection{Sample characterization}
The physical properties of FeGe are highly sensitive to its growth and annealing protocols, which can result in samples with long-range, short-range, or suppressed CDWs \cite{wu_annealing-tunable_2024,shi_annealing-induced_2024,chen_discovery_2024,Klemm2025}. Systematic annealing studies showed that $T_{\rm CDW}$ and $T_{\rm IC}$ are positively correlated, and are both maximized in FeGe samples annealed at 320~$^{\circ}$C, which exhibits a long-range CDW.  

The FeGe samples used in this study were characterized using specific heat and magnetic susceptibility measurements, as shown in Figs.~\ref{Fig_basic}(d) and (e). A large jump is seen in the specific heat $C_{p}$ at $T_{\rm CDW}\approx115$~K, evidencing a first-order transition into the long-range CDW state, and no clear $C_p$ anomaly is observed at lower temperatures. For comparison, in samples with a short-range CDW or no CDW, the $C_p$ peak is much smaller or absent \cite{teng_discovery_2022,shi_annealing-induced_2024}. The magnetic susceptibility $\chi_c$ (measured under a 1~T $c$-axis field) shows a clear upturn below $T_{\rm IC}\approx65$~K, with a much smaller drop at $T_{\rm CDW}$. These behaviors are consistent with previous reports of FeGe with a long-range CDW \cite{Ma2025}.

\begin{figure}
	\includegraphics[width=1\columnwidth]{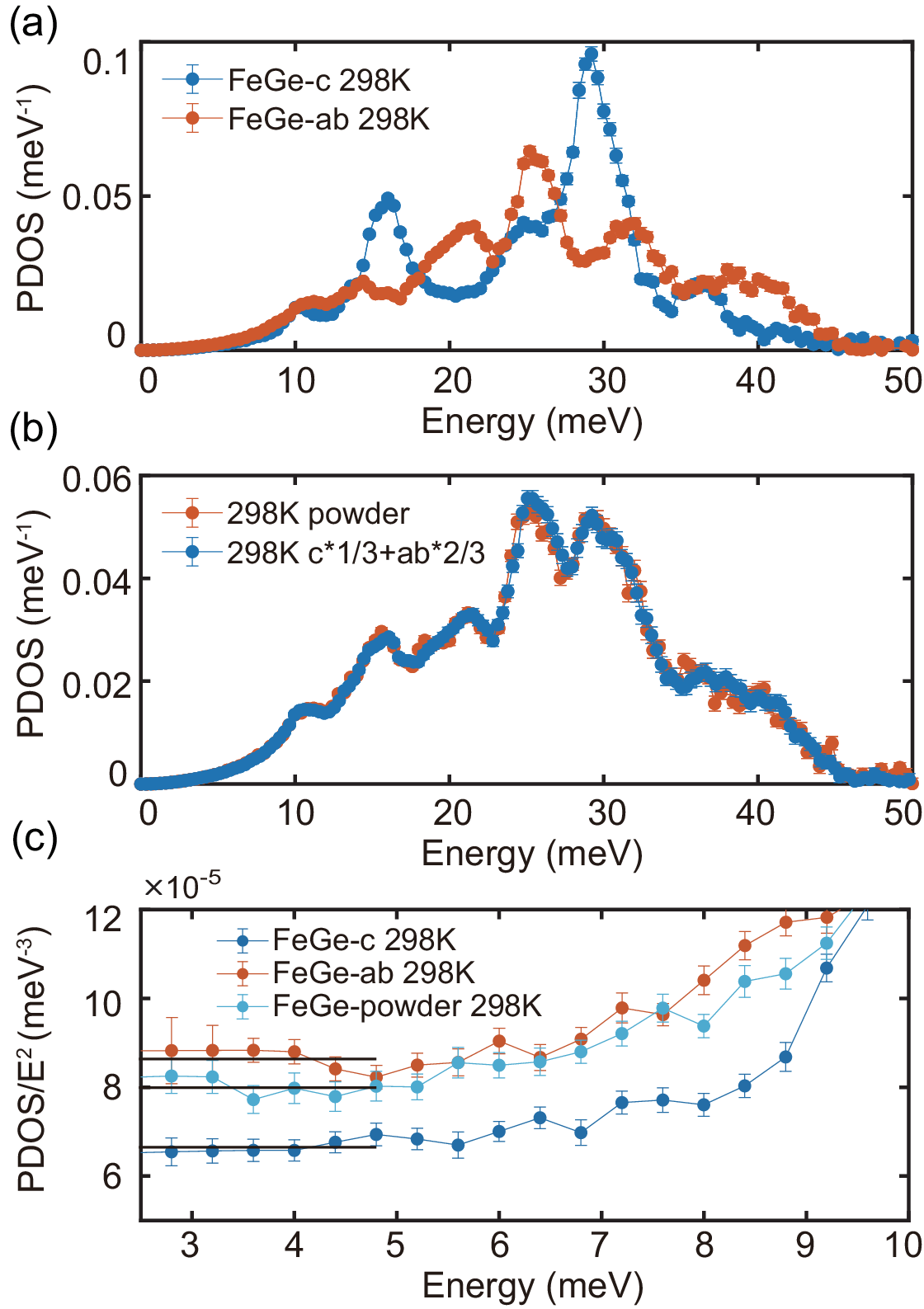} \protect\caption{ (a) The $c$-axis- and $ab$-plane-polarized Fe partial PDOS of FeGe at RT, $g_c(E)$ and $g_{ab}(E)$. (b) Comparison of the Fe partial PDOS $g(E)$, with $\frac{1}{3}g_c(E)+\frac{2}{3}g_{ab}(E)$. (c) Comparison of $g(E)/E^2$, $g_{c}(E)/E^2$, and $g_{ab}(E)/E^2$. The solid lines represent the fit Debye levels.}
	\label{PDOS_RT}
\end{figure}

\begin{figure}
	\includegraphics[width=1\columnwidth]{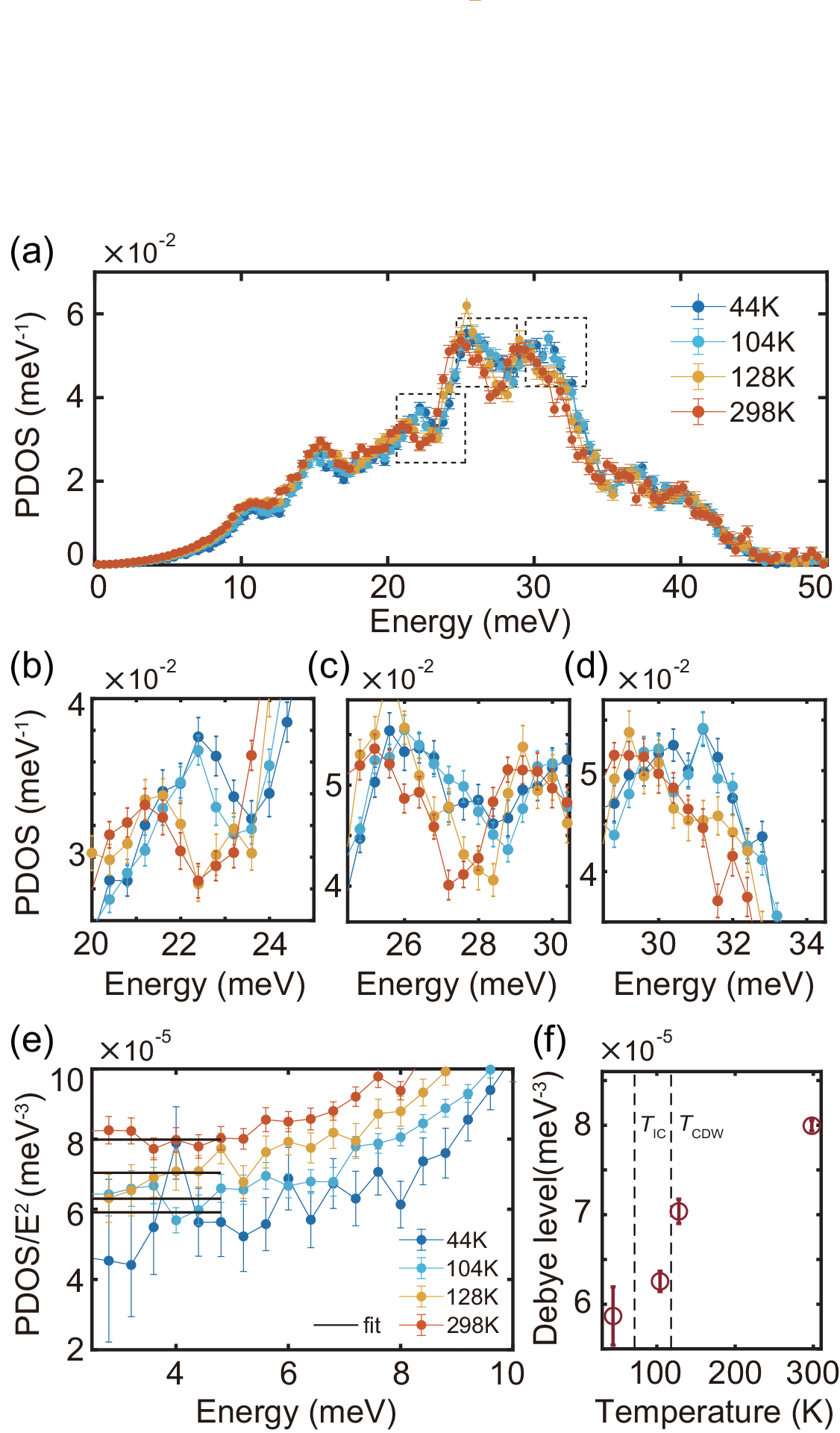} \protect\caption{(a) The Fe partial PDOS of FeGe $g(E)$, measured at several temperatures. Zoomed-in views of $g(E)$ are shown around (b) $\sim22$~meV, (c) $\sim27$~meV, and (d) $\sim31$~meV. (d) Temperature dependence of $g(E)/E^2$. The solids lines are fits to obtain the Debye level. (f) Temperature dependence of the Debye level.}
	\label{PDOS_Tdep}
\end{figure}

\subsection{Nuclear inelastic scattering measurements}

The Fe partial PDOS measurements of FeGe at 298~K (RT) are summarized in Fig.~\ref{PDOS_RT}. By aligning the incident X-rays with a particular crystal axis, NIS measurements selectively probe the partial PDOS polarized along this axis. Fig.~\ref{PDOS_RT}(a) compares the $c$-axis  and $ab$-plane polarized PDOS [$g_{ab}(E)$ and $g_c(E)$], revealing peaks at 16~meV and 29~meV with a dominant $c$-axis polarization, and peaks at 21~meV, 25~meV, 32~meV, and 40~meV with a dominant $ab$-plane polarization. 

Fig.~\ref{PDOS_RT}(b) compares the powder averaged Fe partial PDOS $g(E)$ of FeGe obtained from powder measurements, and the weighted average of the $c$-axis and $ab$-plane polarized PDOS in Fig.~\ref{PDOS_RT}(a). As can be seen, the two agree remarkably well. Because only one in-plane orientation was experimentally measured, this agreement suggests that the Fe partial PDOS of FeGe is essentially isotropic in the $ab$-plane. 

For acoustic phonons, the Debye level $\alpha=\lim_{E\to0}g(E)/E^2$ probes the mean sound velocity via:
\begin{equation*}
\lim_{E\to0}\frac{g(E)}{E^2}=\eta\frac{1}{2\pi^2\hbar^3n\langle v\rangle^3},
\end{equation*}
where $n$ is the number of atoms per unit volume, $\langle v\rangle$ is the mean sound velocity, and $\eta=\frac{2m}{m+M}$ \cite{Chumakov2009,Ksenofontov2010}. Here $m\approx56.94$ for Fe and $M\approx72.63$ for Ge. For FeGe powder at RT, extrapolation of $\frac{g(E)}{E^2}$ to zero energy yields $\alpha=(7.99\pm0.06)\times 10^{-5}$~$\mathrm{meV}^{-3}$ [Fig.~\ref{PDOS_RT}(c)], which in turn gives $\langle v \rangle=(3.06\pm0.01) $~km/s. For comparison, the Debye levels for $c$-axis and $ab$-polarized PDOS are $\alpha_c= (6.67\pm0.04)\times 10^{-5}$~$\mathrm{meV}^{-3}$ and $\alpha_{ab}= (8.59\pm0.07)\times 10^{-5}$~$\mathrm{meV}^{-3}$ [Fig.~\ref{PDOS_RT}(c)], giving $\langle v_{c} \rangle=(3.25\pm0.01)$~km/s and $\langle v_{ab} \rangle=(2.98\pm0.01)$~km/s. Note that within experimental uncertainties, these values fulfill the relationship $\frac{1}{3}\alpha_c+\frac{2}{3}\alpha_{ab}=\alpha$, consistent with the overall agreement of PDOS in Fig.~\ref{PDOS_RT}(b). 

To investigate the evolution of the Fe partial PDOS across $T_{\rm CDW}$ and $T_{\rm IC}$, NIS measurements were performed on FeGe powders at several temperatures, with results summarized in Fig.~\ref{PDOS_Tdep}. The overall features of the PDOS are not strongly affected by the transitions at $T_{\rm CDW}$ and $T_{\rm IC}$, although optical modes around 22~meV, 27~meV, and 31~meV clearly harden upon entering the CDW phase [Fig.~\ref{PDOS_Tdep}(a)]. For comparison, the optical modes at 16~meV and 40~meV do no change significantly in energy across $T_{\rm CDW}$. Zooming into the modes at 22~meV, 27~meV, and 31~meV reveals that the hardening occurs within a small temperature range crossing $T_{\rm CDW}$, whereas no significant changes are seen across $T_{\rm IC}$ [Figs.~\ref{PDOS_Tdep}(b)-(d)]. 

Accompanying the hardening of optical modes across $T_{\rm CDW}$, a hardening of the acoustic phonons is also observed [Fig.~\ref{PDOS_Tdep}(e) ], as can also be seen in the temperature dependence of the Debye level [Fig.~\ref{PDOS_Tdep}(f)]. The hardening of acoustic phonons is unlikely to originate purely from a thermal contraction of the lattice that typically hardens all phonons in the PDOS, which is not experimentally observed. The decrease of Debye level at 128~K relative to RT is consistent with an absence of soft phonons in the CDW formation of FeGe \cite{miao_signature_2023,Subires2025}, whereas soft phonons are observed in kagome metals $A$V$_3$Sb$_5$ \cite{K135, Cs135}, ScV$_6$Sn$_6$ \cite{ScV6Sn6_Cao, ScV6Sn6_Kor}, and LuNb$_6$Sn$_6$ \cite{LuNb6Sn6}. It should be noted that because our measurements probe the Fe partial PDOS, they do not rule out soft phonons associated with Ge. 


\subsection{First-principles phonon calculations }
\begin{figure}
	\includegraphics[width=1\columnwidth]{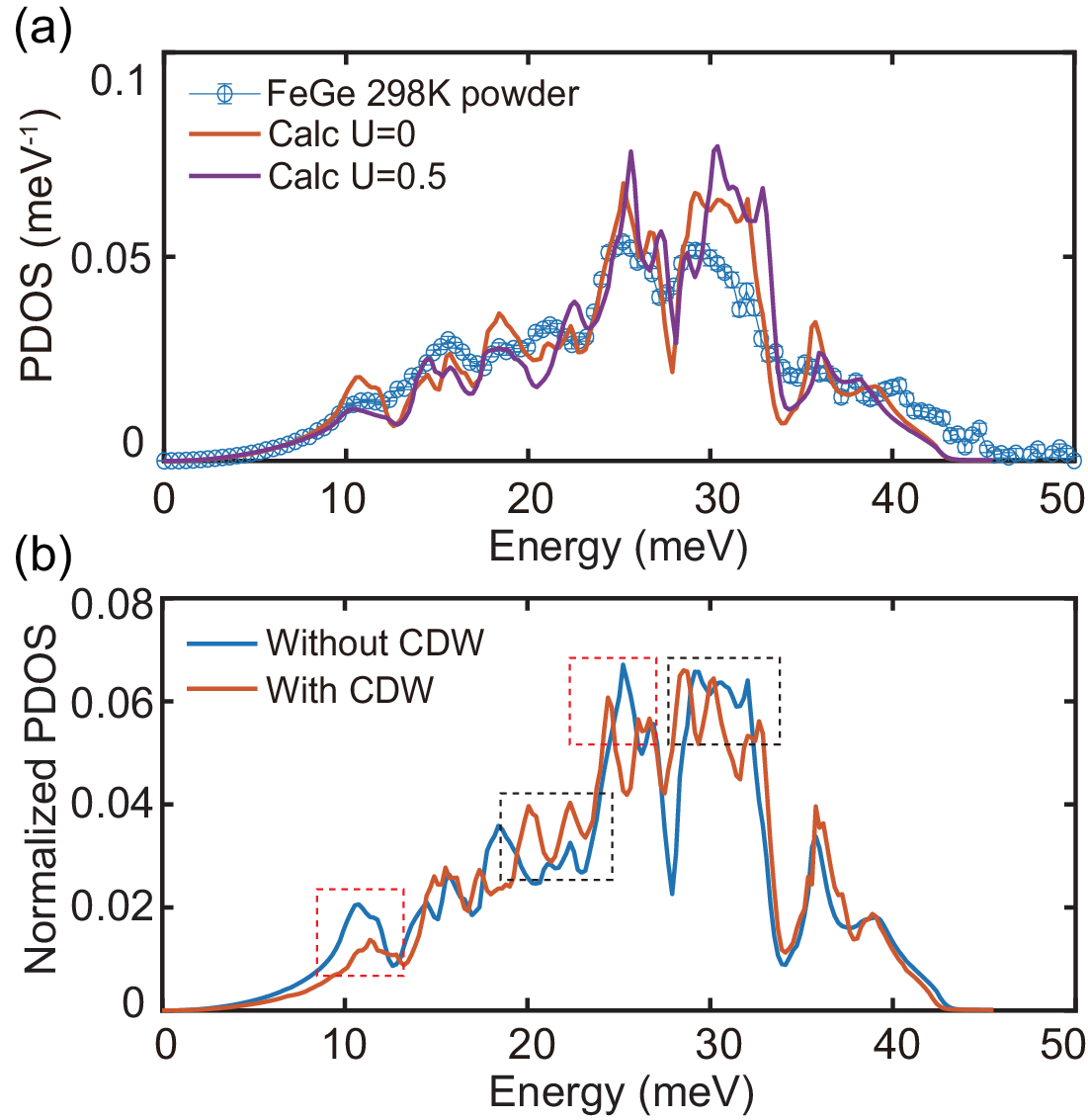} \protect\caption{(a) Calculated Fe partial PDOS with and without Coulomb interaction $U$. (b) Comparison of the calculated Fe partial PDOS of FeGe with and without CDW. The dashed boxes highlight energies ranges that are significantly by the CDW.}
	\label{PDOS_DFT}
\end{figure}

To understand the evolution of lattice dynamics across $T_{\rm CDW}$, the Fe partial PDOS were calculated using density functional theory, with results summarized in Fig.~\ref{PDOS_DFT}. Fig.~\ref{PDOS_DFT}(a) shows the calculated PDOS for FeGe without CDW, showing an overall reasonable agreement with experimental data at RT. DFT underestimates the PDOS above $\sim40$~meV and overestimates the PDOS around $\sim31$~meV, which is unaffected by the soft force constant and the Coulomb interaction $U$ in the calculations. Such a systematic deviation between DFT and experimental PDOS may result from the presence of significant electronic correlations in FeGe, which cannot be captured using single-particle calculations.  

Fig.~\ref{PDOS_DFT}(b) compares the calculated Fe partial PDOS of FeGe with and without the CDW. Several outstanding differences are highlighted in the dashed boxes, including the hardening of acoustic phonons, and changes to the PDOS roughly around 21~meV, 25~meV, and 30~meV. Notably, the high-energy PDOS for $E\gtrsim35$~meV does not change significantly. These changes are in qualitative agreement with the experimental results in Fig.~\ref{PDOS_Tdep}. However, because the changes to the calculated PDOS is rather complex, it is not straightly to determine whether a particular optical mode hardens or softens. Nonetheless, as shown by the dashed boxes in Figs.~\ref{PDOS_Tdep} and \ref{PDOS_DFT}, regarding the energies where there is significant changes to the PDOS, the agreement between experimental results and calculations is reasonable.   

\subsection{Time-domain M\"{o}ssbauer measurements}

\begin{figure}
	\includegraphics[width=1\columnwidth]{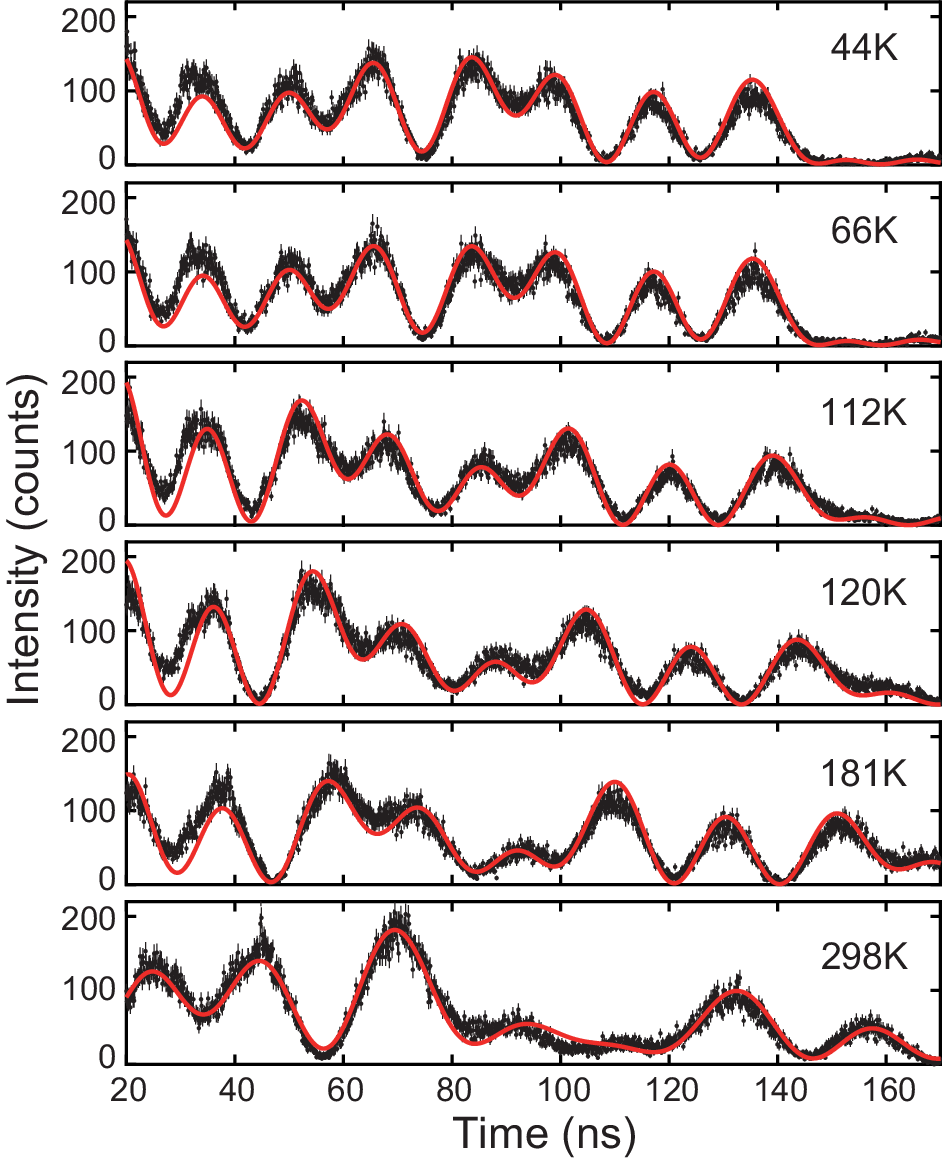} \protect\caption{Time-domain M\"{o}ssbauer spectra of FeGe at representative temperatures. The solid red lines are fits to the model described in the text.}
	\label{Fig_Mossbauer}
\end{figure}

To probe the evolution of magnetic order in annealed FeGe across $T_{\rm CDW}$ and $T_{\rm IC}$, time-domain $^{57}$Fe M\"ossbauer measurements were carried out using synchrotron radiation \cite{Seto2009} on FeGe powder, for temperatures ranging from 44~K to RT, with results shown in Fig.~\ref{Fig_Mossbauer}. Compared to conventional M\"{o}ssbauer spectroscopy in the frequency domain, time-domain measurements are typically faster and allow for a higher signal-to-noise ratio \cite{Seto2013,Yoshida2021}. In the case of FeGe, previous M\"{o}ssbauer studies examined samples without the CDW \cite{tomiyoshi_mossbauer_1966,LennartHaggstrom1975}. 

\begin{figure}
	\includegraphics[width=1\columnwidth]{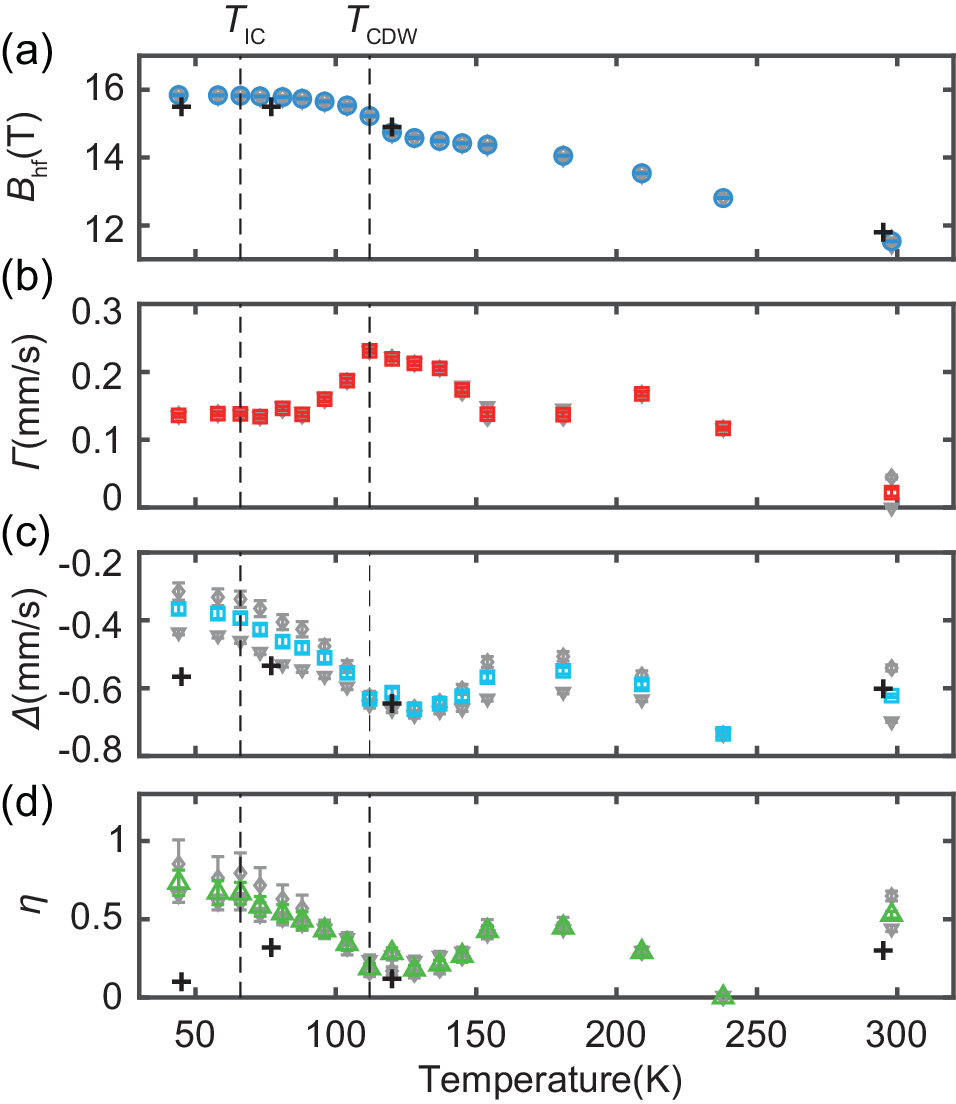} \protect\caption{Temperature dependence of (a) the hyperfine magnetic field $B_{\rm hf}$, (b) the spectral broadening $\Gamma$, (c) the quadrupole interaction energy $\Delta$, and (d) the asymmetry parameter $\eta$. The gray triangles and diamonds represent the fitting results obtained by increasing and decreasing the fixed value $\phi$ by 10 degrees, respectively. The black cross symbols are data from a previous M\"ossbauer study on FeGe without CDW \cite{LennartHaggstrom1975}.}
	\label{Fig_hyperfine_parameters}
\end{figure}

The measured time spectra at various temperatures are shown in Fig.~\ref{Fig_Mossbauer}, and the solid lines are fits obtained using the \texttt{Nexus} software \cite{nexus2024}. The time spectra are modeled by considering a quadrupole splitting, a magnetic hyperfine field, and an additional spectral broadening. 
Note that the isomer shift cannot be extracted from the conducted time-domain M\"{o}ssbauer measurements, as measurements of a reference foil were not performed. 

The quadrupole splitting is determined by the electric field gradient (EFG) tensor, which can be characterized by the three components along its principal axes, with the convention $|V_{zz}|>|V_{yy}|>|V_{xx}|$. Because the EFG tensor is traceless, $V_{xx}+V_{yy}+V_{zz}=0$, there are two independent degrees of freedom, which can be parameterized via the quadrupole interaction energy $\Delta=eQV_{zz}/2$, and the asymmetry parameter $\eta=(V_{xx}-V_{yy})/V_{zz}$. Here, $Q$ is the quadrupole moment of the $I=3/2$ excited state of $^{57}$Fe. The resulting quadruple splitting is $\Delta E_Q=\Delta(1+\eta^2/3)^{1/2}$. The orientation of the EFG tensor is modeled by an isotropic orientational average over all possible EFG orientations, since the experiment is performed on a powder sample. The magnetic hyperfine field $\vec{B}_{\rm hf}=B_{\rm hf}(\theta,\phi)$ experienced by $^{57}$Fe nuclei arises mainly from the spin-polarized surrounding electrons. $\vec{B}_{\rm hf}$ is defined relative to the principal axes of the EFG, with $\theta$ being the angle between $\vec{B}_{\rm hf}$ and $V_{zz}$, while $\phi$ being the angle between the projection of $\vec{B}_{\rm hf}$ onto the $V_{xx}V_{yy}$-plane and the $V_{xx}$ axis, following the convention in previous M\"{o}ssbauer studies \cite{LennartHaggstrom1975}. Finally, an additional broadening of the spectral lines beyond the natural linewidth is accounted for by introducing a Gaussian broadening function with a full width at half maximum of $\Gamma$.


When allowed to vary freely, we were unable to obtain reliable fits for the parameters $\theta$ and $\phi$. We therefore used $\phi=69.0^\circ$ reported previously \cite{LennartHaggstrom1975} and obtained $\theta=86.6(1.6)^{\circ}$, when averaged over temperatures. We note the fit results are overall insensitive to the choice of $\phi$, as shown by fit results with $\phi=69\pm10^\circ$ [gray symbols in Fig.~\ref{Fig_hyperfine_parameters}]. Given the ordered moments of FeGe for $T>T_{\rm IC}$ are oriented along the $c$-axis, the value of $\theta$ suggests that $V_{yy}$ is approximately along the $c$-axis, and $V_{xx}$ and $V_{zz}$ are approximately in the $ab$-plane. 

The values of $\theta$ and $\phi$ were then fixed, and the time domain spectra were fit to obtain $B_{\rm hf}$, $\mathit{\Delta}$, $\eta$, and $\Gamma$. The fitted curved are shown as solid lines in Fig.~\ref{Fig_Mossbauer}, and the fitted parameters are summarized in Fig.~\ref{Fig_hyperfine_parameters}. For comparison, the hyperfine parameters from Ref.~\cite{LennartHaggstrom1975} are indicated by the black cross symbols. Upon cooling through $T_{\rm CDW}$, $B_{\rm hf}$ exhibits a pronounced enhancement [Fig.~\ref{Fig_hyperfine_parameters}(a)], consistent with neutron scattering measurements that showed AFM order being enhanced by the CDW \cite{teng_discovery_2022, Klemm2025}. Concurrently, $\mathit{\Delta}$ and $\eta$ are also enhanced below $T_{\rm CDW}$ [Figs.~\ref{Fig_hyperfine_parameters}(c) and (d)], indicating that the CDW also strongly modifies the EFG at the Fe sites. $\Gamma$ reaches a maximum at $T_{\rm CDW}$ and decreases below this temperature [Fig.~\ref{Fig_hyperfine_parameters}(b)]. 

Furthermore, $\mathit{\Delta}$ and $\eta$ display broad maxima centered around approximately 170~K. This observation may be related to a broad anomaly between 150~K-250~K in samples annealed for less than 8 days~\cite{wu_annealing-tunable_2024}, as our samples were annealed for 96~hours. Annealing was found to affect the CDW via the distribution of Ge vacancies in FeGe, with a uniform distribution detrimental for long-range CDW, while the segregation of Ge vacancies allows the formation of CDW in stoichiometric FeGe \cite{Klemm2025}. This suggests that in addition to distinct Fe sites in the CDW phase of FeGe \cite{chen_discovery_2024, Shi2025a}, there are likely distinct phases of FeGe (with little or concentrated Ge vacancies) probed in our measurements. Because only a single Fe site is considered in our analysis, the fit parameters should be viewed as averaged values across Fe sites. The 170~K broad peaks in $\Delta$ and $\eta$ likely result from such an averaging, with different sites evolving differently in temperature. This view is consistent with the temperature dependence of $\Gamma$, which mirrors those of $\Delta$ and $\eta$, showing a broad dip around 170~K.

\begin{figure}
	\includegraphics[width=1\columnwidth]{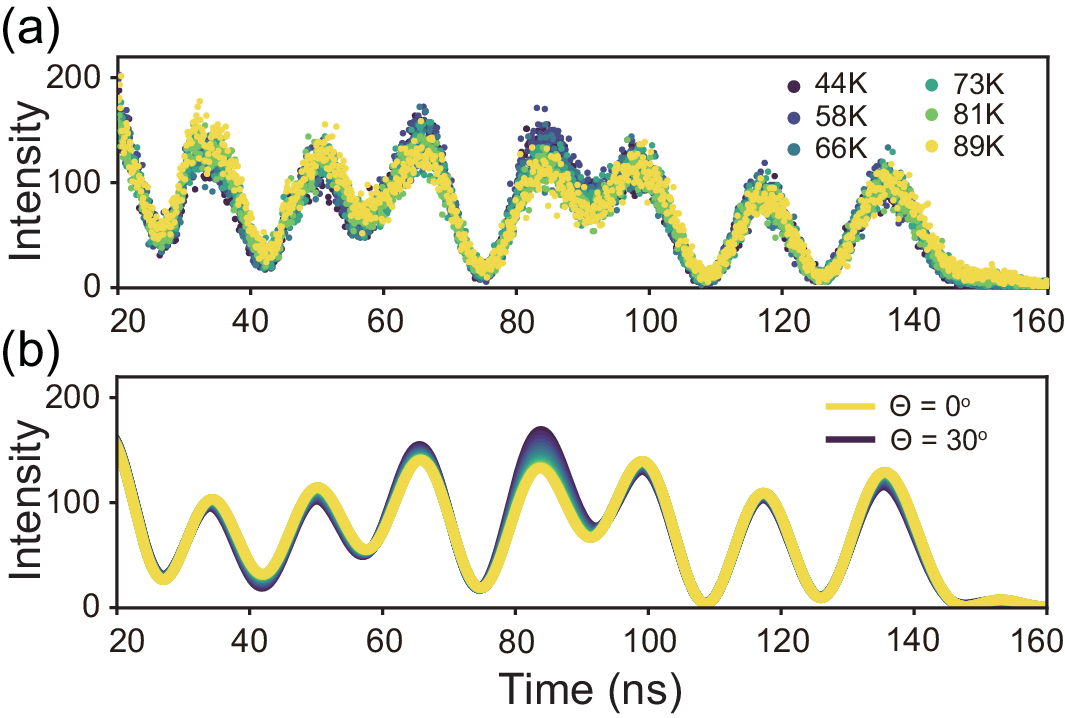} 
    \protect\caption{(a) Experimental time-domain M\"ossbauer spectra of FeGe at temperatures from 44~K to 89~K. (b) Calculated time-domain M\"ossbauer spectra based on a model with the magnetic hyperfine field $\vec{B}_{\rm hf}$ distributed over a cone with an opening angle~$\Theta$, with $\Theta$ ranging from $0^\circ$ to $30^\circ$.}
	\label{Fig_spin_cone}
\end{figure}

To examine how the IC magnetic transition affects the time-domain M\"ossbauer spectra, the measured spectra below 90~K are directly compared in Fig.~\ref{Fig_spin_cone}(a), with prominent enhancements upon cooling around 65~ns and 85~ns. To understand the origin of these effects,  we extended the model in Fig.~\ref{Fig_Mossbauer} by setting the direction of $\vec{B}_{\rm hf}$ to be an angle $\Theta/2$ from its original direction (along $c$-axis), and trace out a cone with an opening angle of $\Theta$ when averaged over the sample [Fig.~\ref{Fig_basic}(a)]. By fixing the magnitude of $\vec{B}_{\rm hf}$ and varying $\Theta$, the simulated time-domain M\"ossbauer spectra are shown in Fig.~\ref{Fig_spin_cone}(b), which captures the experimentally observed enhancement around 65~ns and 85~ns by an increase in $\Theta$. 

Because the oscillation frequency of the M\"ossbauer spectra does not change in temperature, our experimental data indicate that the magnitude of $\vec{B}_{\rm hf}$ does not change upon cooling below $T_{\rm IC}$, whereas our modeling in Fig.~\ref{Fig_spin_cone}(b) suggests the gradual appearance of an in-plane ordered component. Interestingly, such a change already starts to occur above $T_{\rm IC}$ [Fig.~\ref{Fig_spin_cone}(a)], which likely results from slow fluctuations of the IC magnetic order, which were observed in inelastic neutron scattering measurements \cite{chen_competing_2024,Klemm2025}.

The results in Fig.~\ref{Fig_spin_cone} show that although the in-plane and $c$-axis components of the IC order have distinct origins (itinerant and local-moment, respectively) \cite{chen_competing_2024,Klemm2025}, their combined effects, as seen by a local probe, are reasonably described by a spin cone structure with a fixed moment size [Fig.~\ref{Fig_basic}(a)]. 



\section{Conclusion}
The Fe partial PDOS and time-domain M\"{o}ssbauer measurements were performed for annealed kagome FeGe with long-range CDW. Significant hardening of acoustic phonons and optical phonons are observed across the CDW transition, whereas much smaller changes to the PDOS occur across the IC magnetic transition. Time-domain M\"{o}ssbauer measurements evidence an enhancement of the hyperfine field in the CDW state, consistent with neutron scattering measurements. The hyperfine field is found to remain unchanged across the IC transition, with the time-domain M\"ossbauer spectra consistent with a double cone magnetic structure that has both in-plane and $c$-axis ordered moments. These results provide a local-probe perspective on the interplay between and CDW and magnetism in FeGe, evidencing an intertwined nature of lattice, charge, and spin degrees of freedom in kagome FeGe.



\section*{Acknowledgments}

The work at Zhejiang University was supported by the National Key R\&D Program of China (No. 2022YFA1402200), the National Natural Science Foundation of China (No. 12350710785, 12274363), and the Fundamental Research Funds for the Central Universities (Grant No. 226-2024-00068). Yilin Wang was supported by the National Key R\&D Program of China (No.~2023YFA1406304), the Quantum Science and Technology-National Science and Technology Major Project (No. 2021ZD0302803) and the New Cornerstone Science Foundation. 
Single crystal growth and characterization efforts at Rice University are supported by U.S. NSF Award No. DMR-2401084 and the Robert A. Welch Foundation under Grant No. C-1839, respectively (P.D.).
We acknowledge Deutsches Elektronen-Synchrotron (DESY), a member of the Helmholtz Association (HGF), for the provision of experimental facilities. Parts of this research were carried out at the beamline P01 at PETRA~III under proposal I-20231311.

\bibliographystyle{apsrev4-1}
\bibliography{bibfile}
\end{document}